\documentclass[%
 aip,
amsmath,amssymb,
 reprint,%
]{revtex4-1}

\usepackage{graphicx}% Include figure files
\usepackage{dcolumn}% Align table columns on decimal point
\usepackage{bm}% bold math
\usepackage{cleveref}
\usepackage[utf8]{inputenc}
\usepackage[T1]{fontenc}
\usepackage{mathptmx}
\usepackage{etoolbox}
\usepackage{xcolor}

\makeatletter
\def\@email#1#2{%
 \endgroup
 \patchcmd{\titleblock@produce}
  {\frontmatter@RRAPformat}
  {\frontmatter@RRAPformat{\produce@RRAP{*#1\href{mailto:#2}{#2}}}\frontmatter@RRAPformat}
  {}{}
}%
\makeatother
\begin{document}

\newcommand*\mycommand[1]{\texttt{\emph{#1}}}
\graphicspath{{figures/}}

\title{Real-time dynamics with bead-Fourier path integrals. II. Bead-Fourier RPMD}

\author{Nathan London}
\affiliation{Division of Energy, Matter and Systems, School of Science and Engineering, University of Missouri $-$ Kansas City, Kansas City 64110, MO United States.}
\author{Mohammad R. Momeni}
\email{mmomenitaheri@umkc.edu}
\date{}
\affiliation{Division of Energy, Matter and Systems, School of Science and Engineering, University of Missouri $-$ Kansas City, Kansas City 64110, MO United States.}

\date{\today}% It is always \today, today,

\begin{abstract}
Feynman path integrals (PIs) have found many uses in approximate quantum dynamics methods for efficiently calculating real-time quantum correlation functions. The PIs typically take the form of discrete imaginary-time slices along a closed path, where the slices form the ``beads'' of a ring polymer (RP) necklace. Some methods, such as centroid molecular dynamics (CMD), use the RP to generate an effective potential for the dynamics, while others, like RP molecular dynamics (RPMD), directly utilize the RP in real-time dynamics in order to incorporate approximate quantum effects. The standard, discretized bead forms of CMD and RPMD may require a large number of RP beads to obtain accurate results for systems with strong quantum effects and at low temperatures. In Paper I\cite{octLondon2025}, we introduced the bead-Fourier (BF)-CMD method, where we utilized the inclusion of a Fourier sine series to reduce the number of beads needed to converge the CMD effective potential up to eightfold. In this work, we extend RPMD to incorporate BF-PIs in the form of BF-RPMD. We study several implementations of the BF-RPMD method through the calculation of correlation functions for both linear and nonlinear operators. We found that the effectiveness of this method is sensitive to both the type of system and the form of the operators being studied. Our results show that this method is able to produce results on par with standard RPMD, with at worst twofold and up to eightfold reduction in the number of beads by including two to three Fourier components.

\end{abstract}

\maketitle
\section{Introduction}\label{sec:intro}
%Linear and non-linear TCFs and the importance of NQEs and exact is difficult
Real-time time correlation functions (TCFs) are among the most fundamental tools in statistical mechanics for describing the time evolution of physical observables at or near equilibrium conditions within the linear response theory.\cite{chandler1987introduction} The calculation of TCFs of complex condensed-phase and interface systems with thousands of degrees of freedom (DOFs) has been conveniently performed by solving the equations of motion (EOMs) classically, often using analytical potentials. However, to achieve a predictive level of accuracy, the newly developed theories must also incorporate the essential nuclear quantum effects (NQEs), such as zero-point energy (ZPE) and nuclear tunneling, especially important at low temperatures and for light nuclei. Considering the large scale of these systems, propagating nuclear wavepackets according to the time-dependent Schr\"{o}dinger equation is beyond current computational capabilities. Methods that can efficiently and accurately incorporate NQEs into classical molecular dynamics (MD) simulations are thus a key area of developmental focus.\cite{Markland2018} One particular branch of research is utilizing Feynman's path-integral (PI) formulation of quantum statistical mechanics.~\cite{julFeynman2010}

%Bead PIs
In the most conventionally employed ``bead'' formulation of the PIs (B-PIs), NQEs are captured using Feynman’s imaginary-time path-integrals, where each nuclear DOF is represented by a ring polymer (RP) comprised of $n$ copies, or beads, connected by harmonic springs. The extended phase space of this RP captures NQEs, including ZPE and tunneling, with the exception of quantum interference and coherence effects. For calculating real-time observables, a myriad of approximate methods based on imaginary-time ``bead'' path integral MD (B-PIMD) have been introduced. They include bead centroid MD (B-CMD),\cite{cmd1, cmd2, cmd3, cmd4, cmd5, augJang1999} and bead ring polymer MD (B-RPMD) methods,\cite{augCraig2004, jcp_122_084106, jcp_123_034102} as well as a myriad of more sophisticated methods branched from them,\cite{arpc_64_387, rossi_trpmd, augTrenins2019, novTrenins2022, decFletcher2021, octLawrence2023,janLimbu2025} with some being implemented into software packages for general PI simulations.\cite{augLitman2024,dlq2.0,dlq2.1} However, difficulties are known to arise in systems containing light nuclei at low temperatures, where RPs with a large number of beads are required to converge these simulations.\cite{sepRyu2022}. In addition to the increase in the cost of PI simulations with increasing the number of beads, overly extended RPs are known to lead to non-ergodicity problems in sampling.\cite{sepKorol2019,sepRyu2022} As such, it is highly beneficial to develop efficient PI methodologies that require fewer beads to converge without sacrificing accuracy. 

%Introduce BF-PIs and BF-CMD in Paper I
In Paper I\cite{octLondon2025}, we introduced the alternative bead-Fourier (BF) approach for calculating quantum TCFs involving linear operators in the context of the CMD method. In the new BF-CMD method, the effective centroid potential is calculated using BF-PIs rather than the typical B-PIs. We demonstrated the efficiency and accuracy of this new methodology for calculating linear position autocorrelation functions in 1D harmonic, mildly anharmonic, and quartic systems. We showed that at low temperatures, one can achieve a 4- to 8-fold reduction in the number of beads by adding a single Fourier component to the mildly anharmonic and quartic systems.\cite{octLondon2025} 

%Simplicity of RPMD and CMD not good for non-linear ops
While CMD-based methods can provide accurate TCFs for linear operators, as demonstrated in Paper I,\cite{octLondon2025} their accuracy diminishes quickly for non-linear operators.\cite{novPaesani2008} Accuracy can be improved through several methods. One is to relate the CMD and the exact correlation function by mixing classical and semiclassical centroid representations of the operators.\cite{julReichman2000} Another is to use the CMD spectral density from the TCFs as the base model to perform a maximum entropy analytic continuation (MAEC)\cite{decCaffarel1992,decGallicchio1994,junBonca1996} calculation to generate the final TCFs.\cite{novPaesani2008} Both methods have their own computational complexities that can limit their applicability for large systems.

%BF-RPMD
An alternative approach for utilizing PIs for approximate real-time dynamics, as mentioned previously, is B-RPMD. B-RPMD simply takes the B-PIMD Hamiltonian and uses it for real-time dynamics by setting the fictitious Parrinello-Rahman masses in the kinetic energy term to be the physical mass of the particle and removing the thermostatting (i.e., performs dynamics in the microcanonical NVE ensemble).\cite{augCraig2004} This approach makes the method readily applicable to complex, atomistic systems. B-RPMD has been proven successful in calculating various dynamical properties, including reaction rates\cite{jcp_122_084106,jcp_123_034102}, although with less success in calculating vibrational spectra due to the coupling of the internal modes of the RP to the modes of the physical system, which creates spurious peaks in the spectra.\cite{mayWitt2009,augHabershon2008} B-RPMD also performs better than unmodified B-CMD for non-linear TCFs.\cite{augCraig2004,mayHele2016a}

Based upon our success in using BF-PIs to improve the convergence of CMD with respect to the number of beads with BF-CMD in Paper I,\cite{octLondon2025} as well as the success B-RPMD has seen in a wide number of areas, here we introduce a BF extension of RPMD through a BF-RPMD method. Our motivation for developing a BF-RPMD method to complement our BF-CMD method is threefold. First is the potential applicability to non-linear TCFs for which CMD methods struggle. Second, a BF-RPMD method would be far simpler to implement in large-scale atomistic simulations. As it stands, BF-CMD is limited to simple model systems without the development of an adiabatic or partially adiabatic version of the method. Directly utilizing the BF-PIMD Hamiltonian for real-time dynamics in a BF-RPMD framework provides a natural path to more complex systems. Lastly, a BF-RPMD method is likely to be more computationally efficient than an adiabatic BF-CMD method, where the integration time step would need to be reduced to accommodate the scaled masses that would need to be used to properly sample the effective BF-CMD potential.

%Paper outline
Here, in Paper II, we outline our formulation for the new real-time BF-RPMD method. We showcase the accuracy of this method by calculating linear and non-linear position autocorrelation functions in 1D harmonic, mildly anharmonic, and quartic model systems. By comparing the results of BF-RPMD with its B-RPMD and BF-CMD predecessors, as well as with our exact data, we demonstrate its accuracy and efficiency in calculating real-time properties. This paper is organized as follows: in Section II, following a brief introduction to BF-PIMD, we present our formulation of the new BF-RPMD method and the derivations of the EOMs, presented here for the first time for a general external potential. Simulation details are provided in Section III, followed by results and discussions in Section IV for TCFs involving both linear and non-linear operators. We conclude this paper with conclusions and future work in Section V.

\section{Theory}\label{sec:theory}
  \subsection{Bead-Fourier path integrals}\label{sec:bf-pi}
    Here, we give a brief overview of the BF-PI methodology. A more detailed description can be found in Paper I.\cite{octLondon2025} In the BF-PI framework, the imaginary time paths used to calculate the quantum canonical partition function are described using a number of discrete imaginary time slices (the ``beads'' of standard PIs) that are connected through continuous paths represented as a Fourier sine series,\cite{febVorontsov1997,junIvanov2003}
  \begin{equation}
    q_{j}(\xi) = q_{j} + (q_{j+1}-q_{j})\xi + \sum^{k_{\mathrm{max}}}_{k=1}a_{jk}\sin(k\pi\xi),
    \label{eq:path}
  \end{equation}

  where $q_{j}$ is the position of the $j^{th}$ bead, $a_{jk}$ is the amplitude of the $k^{th}$ Fourier component for the $j^{th}$ bead, and $\xi$ represents the distance along the imaginary time interval between adjacent beads and goes between 0 and 1. The Fourier series is set to only include $k_{\mathrm{max}}$ terms for practical simulations.
   
  The canonical partition function in the BF-PI framework is,\cite{febVorontsov1997,junIvanov2003}
  \begin{equation}
    %Z = \propto \int\mathrm{d}\{q_j\}\int\mathrm{d}\{a_{jk}\}\ \mathrm{e}^{-\beta H(q_j,a_{jk})}.
    Z_{\mathrm{BF}}  \propto \int\mathrm{d}\mathbf{q}\int\mathrm{d}\mathbf{a}\ \mathrm{e}^{-\beta H_{\mathrm{BF}}(\mathbf{q},\mathbf{a})},
  \end{equation}
  where $ \{\mathbf{q}\} $ and $ \{\mathbf{a}\} $ are the set of bead positions and Fourier amplitudes and
  $\beta=(k_{\mathrm{B}}T)^{-1}$ is inverse temperature. The BF-Hamiltonian is,\cite{febVorontsov1997,junIvanov2003}
  \begin{multline}
    %H(q_j,a_{jk})= \sum_{j=1}^{n} \left[ \frac{1}{2}\omega_{n}^2\left((q_{j+1}-q_j)^2 + \sum^{k_\mathrm{max}}_{k=1}
    H_{\mathrm{BF}}(\mathbf{q},\mathbf{a})= \sum_{j=1}^{n} \left[ \frac{1}{2}m_{n}\omega_{n}^2\left((q_{j+1}-q_j)^2 + \sum^{k_\mathrm{max}}_{k=1}
      \frac{(k\pi)^2}{2}a_{jk}^2\right) \right. \\ \left.   + \frac{1}{n}\int_0^1\mathrm{d}\xi\ V(q_j(\xi))\right],
      \label{eq:bf-ham}
  \end{multline}
  with $m_{n} = \frac{m}{n}$, $m$ is the mass of the particle, $\omega_{n} = \frac{n}{\beta\hbar} $ and $V(q)$ is the external potential.
  \subsection{Bead-Fourier Path Integral Molecular Dynamics}\label{sec:bf-pimd}
    %\begin{itemize}
        In order to perform dynamical simulations with BF-PIs, the Hamiltonian in
        Eq.~\ref{eq:bf-ham} must be extended with the corresponding conjugate momenta
        for both the beads and Fourier amplitudes,\cite{junIvanov2003}
        \begin{equation}\label{eq:bf-plus-p}
          H \to H + \sum^{n}_{j=1}\left[ \frac{p_{j}^2}{2m^{'}} + \sum^{k_{\mathrm{max}}}_{k=1}
          \frac{p_{jk}^2}{2m_{k}} \right],
        \end{equation}
        where $ p_{j} $ is the momentum of the $ j^{th}$ bead, $ p_{jk} $ is the conjugate
        momentum of the $k^{th}$ Fourier amplitude for the $j^{th}$ bead, $ m^{'} $  is the fictitious Parrinello-Rahman
        mass of the beads\cite{janParrinello1984a} and $m_{k}$ is the mass
        for that Fourier amplitude.

    To be able to perform dynamical simulations, we need to derive a set of EOMs for the system. The EOMs for the bead-Fourier system can be found as,
    \begin{equation}
      \frac{ \partial p_{j}}{ \partial t} = -\frac{ \partial H}{ \partial q_{j}},
      \label{eq:eom-p-bead}
    \end{equation}
    \begin{equation}
      \frac{ \partial q_{j}}{ \partial t} = \frac{ \partial H}{ \partial p_{j}} = \frac{p_{j}}{m^{'}},
      \label{eq:eom-q-bead}
    \end{equation}
    \begin{equation}
      \frac{ \partial p_{jk}}{ \partial t} = -\frac{ \partial H}{ \partial a_{jk}},
      \label{eq:eom-p-fourier}
    \end{equation}
    and
    \begin{equation}
      \frac{ \partial a_{jk}}{ \partial t} = \frac{ \partial H}{ \partial p_{jk}} = \frac{p_{jk}}{m_{k}}.
      \label{eq:eom-a-fourier}
    \end{equation}
    The derivative of the Hamiltonian with respect to the bead positions is
    \begin{equation}
    \begin{aligned} 
      -\frac{ \partial H}{ \partial q_j} 
      &= -\frac{ \partial }{ \partial q_j} \left[ \sum^{n}_{j=1} \left[\frac{1}{2}m_{n}\omega_n^2
      (q_{j+1}-q_j)^2 
      %\right] + \frac{ \partial }{ \partial x_j} \left[ \sum^{n}_{j=1} 
      +\frac{1}{n} \int_0^1 
  \mathrm{d}\xi\ V[q_j(\xi)] \right] \right] \\
      &= -m_{n}\omega_n^2(2q_j-q_{j+1}-q_{j-1})\\
      &- \frac{1}{n}\left[
      \int_0^1 \mathrm{d}\xi\ \frac{ \partial V[q_j(\xi)]}{ \partial q_j(\xi)} \frac{ \partial q_j(\xi)}{ \partial q_j}
      \right. \\
      &\left.
      +\int_0^1 \mathrm{d}\xi\ \frac{ \partial V[q_{j-1}(\xi)]}{ \partial q_{j-1}(\xi)} \frac{ \partial q_{j-1}(\xi)}{
      \partial q_j}\right]\\
      &= -m_{n}\omega_n^2(2q_j-q_{j+1}-q_{j-1})\\
      &- \frac{1}{n}\left[
      \int_0^1 \mathrm{d}\xi\ \frac{ \partial V[q_j(\xi)]}{ \partial q_j(\xi)}(1-\xi)
      +\int_0^1 \mathrm{d}\xi\ \frac{ \partial V[q_{j-1}(\xi)]}{ \partial q_{j-1}(\xi)}\xi
      \right] .
      \label{eq:part-x}
    \end{aligned}
    \end{equation}

    The derivative of the Hamiltonian with respect to the Fourier amplitudes is
    \begin{equation}
    \begin{aligned}
    -\frac{ \partial H}{ \partial a_{jk}}  
      &= -\frac{1}{2} m_{n}\omega_{n}^{2}(k\pi)^2 a_{jk} - \frac{1}{n}\int_{0}^{1} \mathrm{d} \xi\
    \frac{ \partial V[q_{j}(\xi)]}{ \partial q_{j}(\xi)} \frac{ \partial q_{j}(\xi)}{ \partial a_{jk}}\\
      &= -\frac{1}{2} m_{n}\omega_{n}^{2}(k\pi)^2 a_{jk} - \frac{1}{n}\int_{0}^{1} \mathrm{d} \xi\
    \frac{ \partial V[q_{j}(\xi)]}{ \partial q_{j}(\xi)} \sin(k\pi\xi).
    \label{eq:part-a}
  \end{aligned}
  \end{equation}

Eqs. \ref{eq:part-x} and \ref{eq:part-a} are the first time, to our knowledge, that have been presented for a general external potential. Ref. \citenum{julIvanov2005} does give the force equations for a power-of-distance potentials, but not for general potentials.

A common practice in PIMD/RPMD simulations is to convert from Cartesian coordinates to a set of normal modes such that the RP beads are converted from a set of $n$ coupled harmonic oscillators to $n$ uncoupled harmonic oscillators.\cite{sepCeriotti2010a}
  This removes the need to calculate the forces due to the spring terms and allows for an exact propagation of the free RP terms under harmonic motion.\cite{sepCeriotti2010a} We invoke these RP normal modes for the bead positions as well, where we define
  the BF-PIMD Hamiltonian as,
    \begin{multline}
      H_{\mathrm{BF-PIMD}} = H_{0-\mathrm{bd}}+ \sum_{j=1}^n\sum_{k=1}^{k_{\mathrm{max}}}\left[ \frac{p_{jk}^2}{2m_{k}}+
      \frac{1}{4}m_{n}\omega_{n}^2(k\pi)^2a_{jk}^2 \right]\\  + \frac{1}{n}\sum_{j=1}^n \int_{0}^1 \mathrm{d}\xi\ V(q(\xi)),
    \label{eq:nm-ham}
  \end{multline}
  with the free RP-like term,
  \begin{equation}
    H_{0-\mathrm{bd}} = \sum_{l=0}^{n-1}\left[ \frac{\tilde{p}_l^2}{2m^{'}} +
    \frac{1}{2}m_{n}\omega_{l}^2\tilde{q_{l}}^2\right]
    \label{eq:nm-free}
  \end{equation}
  where $ \tilde{q}_l $ and $ \tilde{p}_l $ are the position and momentum of the $l^{th}$ normal mode, and the normal mode frequencies are given as
    \begin{equation}
      \omega_{l} = 2\omega_{n}\sin\left(\frac{l\pi}{n}\right).
      \label{eq:nm-freq}
    \end{equation}
  The transformation between the Cartesian coordinates and normal modes of the RP is well-defined and simple to perform.

  We perform dynamics under a velocity Verlet (VV) scheme.\cite{julVerlet1967} For the bead positions and momenta, we follow a similar methodology as laid out in Ref. \citenum{sepCeriotti2010a}, but we perform the normal mode transformation on the forces from Eq.~\ref{eq:part-x} (without the first term representing the spring forces) and update the normal mode bead momenta. For the Fourier components, we also do a separation of the harmonic term (the first term in Eq.~\ref{eq:part-a}), which is propagated exactly in the same fashion as the bead positions and the forces from the external potential (the second term in Eq.~\ref{eq:part-a}).

  %\subsection{Correlation Functions}

  %\subsection{BF-PIMD with PILE thermostat (move before?)}
  %To perform the canonical sampling needed for the initial conditions of the correlation functions, we make use of
  %BF-PIMD. We utilize the same Hamiltonian as in BF-RPMD (check mass details) with the addition of a thermostat to
  %ensure sampling of the canonical ensemble. 
  The final component of BF-PIMD is to introduce a thermostat so that the BF-PI trajectories sample the canonical ensemble. Our implementation here differs from previous BF-PIMD methods that utilized a Nos\'{e}-Hoover chain thermostat\cite{augMartyna1992a} and a center-of-mass thermostatting scheme\cite{junIvanov2003,julIvanov2005} or staging coordinates of the beads.\cite{Ivanov2005} We make use of the simple normal mode definition of the RP beads as described previously, and a thermostat designed for PIs, specifically the PI Langevin equation (PILE)\cite{sepCeriotti2010a} thermostat. Under the PILE thermostat, the momenta of the beads and Fourier components are updated as,
  \begin{equation}
    \tilde{p}_l \leftarrow c_{1,l}\tilde{p}_l + \sqrt{\frac{m_{n}}{\beta}}c_{2,l} \zeta_{l}
    \label{eq:PILE-bd}
  \end{equation}
  and
  \begin{equation}
    p_{jk} \leftarrow c_{1,jk}p_{jk} + \sqrt{\frac{m_{k}}{\beta}}c_{2,jk} \zeta_{jk},
    \label{eq:PILE-Fr}
  \end{equation}
  where $\zeta_{l}$ and $\zeta_{jk}$ are independent random Gaussian numbers with unit variance and
  zero mean.
  The coefficients are defined as,
  \begin{equation}
    c_{1,x} = e^{-(\Delta t/2)\gamma_{x}}
    \label{eq:coef-bd}
  \end{equation}
  and 
  \begin{equation}
    c_{2,x} = \sqrt{1-c_{1,x}^2},
    \label{eq:coef-Fr}
  \end{equation}
  where $\Delta t$ is the simulation time step. The friction coefficients, $\gamma_{x}$, are determined such that they produce the smallest autocorrelation time of the harmonic oscillator Hamiltonian\cite{}. For the beads, the coefficients are,
  \begin{equation}
    \gamma_{l} = 
    \begin{cases}
      1/\tau_{0}, \quad &l=0 \\
      2\omega_{l}, \quad &l>0,
    \end{cases}
    \label{eq:gaml}
  \end{equation}
  where the time constant $\tau_{0}$ is a free parameter that determines the thermostatting of the centroid as its normal mode frequency, $\omega_{0}$, is equal to zero. As there are no zero-frequency modes of the Fourier components, their friction coefficients are, %\textcolor{red}{(check for no scaling) this is for no scaling}
  \begin{equation}
    \gamma_{jk} = \frac{2}{\sqrt{2}}k\pi\omega_{n}
    \label{eq:friction-Fr}
  \end{equation}

  In Paper I,\cite{octLondon2025} we used a form of the BF Hamiltonian in which a scaling parameter,\cite{Ivanov2005}
  \begin{equation}
    \tilde{a}_{jk} = \frac{k\pi}{\sqrt{2}}a_{jk},
    \label{eq:scale}
  \end{equation}
  was used in all places where $a_{jk}$ appears. This scaling makes the Fourier components less dependent on $k$, so they all oscillate on a similar timescale.\cite{Ivanov2005} This scaling results in a different Hamiltonian, not related through a direct transformation, but gives the same results for thermal properties.\cite{Ivanov2005} Here, we have also considered the effects of this scaling, which will be discussed in Sec.~\ref{sec:sampling}.

  \subsection{Bead-Fourier Ring Polymer Molecular Dynamics}\label{sec:bf-rpmd}
  As with conventional RPMD, the extension to BF-RPMD from BF-PIMD is very straightforward. The real-time dynamics is performed directly from the Hamiltonian in Eq.~\ref{eq:nm-ham} with $m^{'}$ set to the physical mass of the particle.
  In our method, we perform both BF-PIMD and BF-RPMD at the physical temperature, so the mass is scaled to $m_{n}$ in our BF-RPMD simulations. Additionally, the dynamics are performed in the microcanonical ensemble with no thermostatting.

  Additionally, the inclusion of the Fourier components will have an effect on the dynamical properties by altering the forces coming from the external potential. Also, their harmonic motion could couple to the external potential, introducing artifacts in the dynamics. To fully assess this, we perform dynamics both with and without thermostatting the Fourier components. We use the same PILE thermostat of Eq.~\ref{eq:PILE-Fr}, and note that we do not thermostat any of the bead normal modes. Further details of the different methods will be discussed in the simulation details.

  \subsection{BF-RPMD Correlation Functions}\label{sec:CFs}
  To understand the BF-RPMD dynamics, we calculate correlation functions (CFs), which can then be used for a number of different dynamical properties. In a similar manner to B-RPMD, we define the BF-RPMD CF as,
  \begin{equation}
    C_{AB}(t) \propto \frac{1}{Z_{\mathrm{BF}}} \int \mathrm{d}\mathbf{q}\
    \int\mathrm{d}\mathbf{a}\int\mathrm{d}\mathbf{p}\ \mathrm{e}^{-\beta
    H_{\mathrm{BF-RPMD}}}A_{\mathrm{BF},x}(0)B_{\mathrm{BF},x}(t),
    \label{eq:BF-CF}
  \end{equation}
  where $\mathbf{p}$ includes both the bead and Fourier momenta.

  The subscript $x$ on the BF-RPMD operators indicates the form of the BF estimator used to calculate the value of the operator. We use two different estimators, bead and continuous, in a similar fashion to BF-PIMD\cite{junIvanov2003} and BF-CMD in Paper I.\cite{octLondon2025}

  The bead-estimated operators only include the bead positions,
  \begin{equation}
    A_{\mathrm{BF,bd}}(t) = \frac{1}{n}\sum_{j=1}^nA(q_{j}(t)),
    \label{eq:op-bd}
  \end{equation}
  where $q_{j}(t)$ is the $j^{th}$ bead position at time $t$ as evolved under the BF-RPMD Hamiltonian.
  The continuous estimator incorporates the full path information,
  \begin{equation}
    A_{\mathrm{BF,cont}}(t) = \frac{1}{n}\sum_{j=1}^n\int_{0}^1 \mathrm{d}\xi\ A(q_{j}(\xi;t)),
    \label{eq:op-cont}
  \end{equation}
  where $q_{j}(\xi;t)$ is the value along the path starting from bead $j$ at time $t$, again as evolved under the BF-RPMD Hamiltonian.

  For the linear position operator, the two versions of the operator reduce to a centroid form. The bead estimator reduces to the bead centroid,
  \begin{equation}
    x_{\mathrm{BF,bd}} = \frac{1}{n} \sum_{j=1}^n q_{j} = Q_{\mathrm{Bd}},
    \label{eq:bd-cent}
  \end{equation}
  while the continuous estimator reduces to the BF centroid, as defined in Paper I,\cite{octLondon2025}
  \begin{equation}
    x_{\mathrm{BF,cont}} = \frac{1}{n} \sum_{j=1}^n \int_{0}^1 \mathrm{d}\xi\ q_{j}(\xi) = Q_{\mathrm{BF}.}
    \label{eq:bf-cent}
  \end{equation}
  
  To study their applicability in different scenarios, we calculate CFs using both the bead and continuous estimators here.

\section{Simulation Details}\label{sec:sim-det}

\subsection{Model Systems}\label{sec:systems}
We consider three one-dimensional model systems, the harmonic oscillator,
  \begin{equation}
    V(x) = \frac{1}{2}x^2,
    \label{eq:harm}
  \end{equation}
  the mildly anharmonic oscillator,
  \begin{equation}
    V(x) = \frac{1}{2}x^2 + \frac{1}{10}x^3 + \frac{1}{100}x^4, 
    \label{eq:mild}
  \end{equation}
  and the quartic oscillator,
  \begin{equation}
    V(x) = \frac{1}{4}x^4.
    \label{eq:quart}
  \end{equation}
  We use $m=\hbar=1$ and two temperatures, $\beta=1$ and $\beta=8$. 
  These models, although being stringent tests for real-time path integral methods, allow for comparison with exact quantum mechanical results.

  \subsection{Canonical Sampling}\label{sec:sampling}
  The initial conditions needed to calculate the correlation functions are taken from BF-PIMD simulations. We consider $n=2$ and $n=4$ at $\beta=1$ and a range of beads from $n=2$ to $n=32$ at $\beta=8$. We also perform simulations using a range of Fourier components from $k_{\mathrm{max}}=0$ (including only the linear path between beads) to $k_{\mathrm{max}}=5$. All integrals over $\xi$ are performed using the trapezoid rule with 20 segments. We take the fictitious bead mass $m^{'}$ to be $m_{n}$ and the Fourier masses to be $m_{k}=m_{n}$ for all simulations. We utilize the scaled Fourier amplitudes for all simulations. For sampling, we run 32 independent trajectories using a simulation time step of 0.01 a.u. and a PILE time constant, $\tau_{0}$ of 50 a.u. Each trajectory is equilibrated for 2500 a.u. and then propagated for an additional 25,000 a.u. with configurations saved every 250 time steps.

  For BF-CMD, we use the same MC sampling algorithm as in Paper I,\cite{octLondon2025} with 32 independent sets of sampling with a decorrelation length of 500 between saved configurations. For each model system, we consider the converged effective potential determined in Paper I.\cite{octLondon2025} The specific details of the number of beads and Fourier components are given as needed in relevant figures and discussion.

B-RPMD results are calculated using the same details as BF-RPMD Method 1, without including any Fourier components and with no linear path in between beads, removing the integrals over $\xi$ in the external potential.
  \subsection{Correlation Functions}\label{sec:dynamics}
  BF-RPMD simulations are performed by evolving trajectories as discussed in Sec.\ref{sec:bf-rpmd} with a time step of 0.01 a.u. Correlation functions are averaged over 32 sets of 31,250 trajectories for a total of $1\times10^6$ trajectories. The correlation functions for both the linear and non-linear operators are calculated from the same trajectories.

  As mentioned in Sec.~\ref{sec:bf-rpmd}, we add a thermostat to the Fourier components for some simulations. We define two real-time dynamics methods: Method A, which has no thermostatting, and Method B, which thermostats the Fourier components using a strongly damped PILE thermostat. The Fourier components remained scaled to be consistent with the BF-PIMD sampling. 

BF-CMD simulations are performed under the Hamiltonian defined in Eq. 24 in Paper I,\cite{octLondon2025} with a time step of 0.001 a.u. B-RPMD dynamics is performed the same as BF-RPMD, but as with the sampling, without any Fourier components or linear path in between the beads.

  \subsection{Exact Results}
  The exact Kubo-transformed autocorrelation function is defined as,
  \begin{equation}
    \tilde{C}_{AA}(t) = \frac{1}{\beta Z} \int_{0}^{\beta} \mathrm{d}\lambda\ \mathrm{Tr}\left[ e^{-(\beta-\lambda)\hat{H}} \hat{A} 
    e^{-\lambda\hat{H}}e^{i\hat{H}t/\hbar}\hat{A} e^{-i\hat{H}t/\hbar}\right].
    \label{eq:kubo-exact}
  \end{equation}
  We calculate the exact results using the split operator Fourier transform (SOFT)
  method\cite{sepFeit1982,janFeit1983,octKosloff1983} in
  an in-house code, to propagate the wavefunction separately in real time and imaginary time. For all systems, we work in the harmonic oscillator eigenfunction basis on a position grid between -10 and 10 a.u. with a spacing of 0.005 a.u. We use the first 32 basis functions for $\beta=1$ and 16 basis functions for $\beta=8$ to achieve converged correlation functions. Real-time propagation is done with $\Delta t=0.01$, and imaginary time propagation is done with $\Delta \beta=0.01$. The integral over $\lambda$ in Eq.~\ref{eq:kubo-exact} is done using the trapezoid rule with the same imaginary time spacing as the imaginary time propagation. 

  \section{Results and Discussion}\label{sec:results}

  \subsection{Harmonic Oscillator}
  The harmonic oscillator is a great testbed for benchmarking the BF-RPMD method, owing to its simple form and readily obtainable exact quantum results. In this section, we first discuss the low-temperature case, $\beta=8$, starting with BF-PIMD results to validate our use of the PILE thermostat. We present our BF-RPMD data next.

  \subsubsection{Equilibrium Properties}
  
  \begin{figure}
  \begin{center}
    \includegraphics[scale=1]{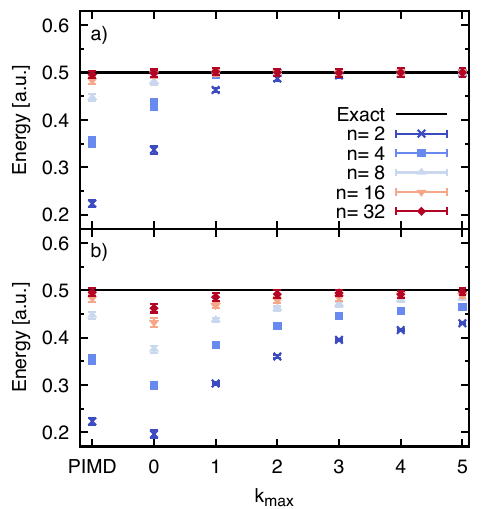}
  \end{center}
  \caption{Quantum energy of the harmonic oscillator of Eq.~\ref{eq:harm} at $\beta=8$ using BF-PIMD with varying numbers of beads and Fourier components, using (a) the bead estimator and (b) the continuous estimator. BF-PIMD with $k_{\mathrm{max}}=0$ includes the linear path between beads, while the first column labeled ``PIMD'' refers to the standard B-PIMD. The exact value is given as the black line.}
  \label{fig:ho-eng}
  \end{figure}
  
% \begin{figure}
 % \begin{center}
  %  \includegraphics[scale=1]{Figures/ho-dist-1.pdf}
 % \end{center}
 % \caption{Position distribution function of the harmonic oscillator (Eq.~\ref{eq:harm}) from BF-RPMD with $n=4$ and $k_{\mathrm{max}}=1 $, comparing the distribution using just the bead positions (blue squares) and integrating over the paths (red diamonds) with the exact quantum distribution (black line).}
 % \label{fig:ho-dist}
 % \end{figure}
  To verify that our version of BF-PIMD is working properly, and to see its connection to BF-RPMD, we look at a few equilibrium properties of the harmonic oscillator of Eq.~\ref{eq:harm} at $\beta=8$. 
  
  The first property we examine is the quantum energy, as shown in Fig.~\ref{fig:ho-eng}. We compare the BF-PIMD energy calculated using the bead estimator in the top panel and the continuous estimator in the bottom panel with the standard B-PIMD energy. The BF-PIMD energies are calculated using the virial estimator, as is done in previous studies.\cite{junIvanov2003} For B-PIMD, it takes 32 beads to reach the exact quantum results. For BF-PIMD, as with the previous studies, the inclusion of the Fourier components reduces the number of beads needed to converge to the exact results, where for $k_{\mathrm{max}}\geq 1$, 4 beads are enough to recover the quantum results when using the bead estimator. Also consistent with previous studies, the convergence of the energy with the continuous estimator is much slower; even with 32 beads, as in standard B-PIMD, several Fourier components are needed to match the exact results.
  
%To better understand the difference between the two estimators, we examine the position probability distributions. Fig.~\ref{fig:ho-dist} shows position distributions for BF-PIMD with $n=4$ and $k_{\mathrm{max}}=1$, which is the converged case for achieving the exact quantum energy using the bead estimator. The figure compares the distributions of the individual bead positions (blue squares) and of the integrals over the paths between beads (red diamonds) with the exact quantum distribution. We see that the bead position distribution matches the exact quantum distribution, while the ``continuous'' position distribution is significantly narrower. The effect of these distributions is reflected in the quantum energy, as the non-linear nature of the energy operator will depend on the distributions of the individual beads/paths. As seen in Fig.~\ref{fig:ho-eng}, the continuously estimated energy for this system with $n=4$ and $k_{\mathrm{max}}=1$ is significantly underestimated, which is consistent with the position distribution.

  Overall, our calculated quantum energies indicate that using the PILE thermostat in our BF-PIMD implementation yields accurate results. In the next section, we discuss the dynamical properties as obtained from the BF-RPMD method for the harmonic oscillator.

  \subsubsection{Dynamical Properties}\label{sec:ho-CF}

The CFs for the harmonic oscillator were calculated using the different BF-RPMD methods presented in Sec.~\ref{sec:dynamics} (i.e., Methods A and B). For all BF-RPMD results, we show converged data with respect to the number of Fourier components (or the maximum of 5 with small numbers of beads). This convergence is determined by both the zero-time limit from the sampling as well as the behavior of the CFs over time.

  \begin{table}
    \caption{Number of Fourier components needed to converge $ \tilde {C}_{xx}(t) $ at $\beta=8$ for the harmonic oscillator using BF-RPMD Methods A and B with a varying number of beads ($n=2$ to $n=32$) for both the bead (Bead) and continuous (Cont.) estimators.}
    \begin{tabular}{ |c | *{3}{ c |}  c |}
    \hline
      & \multicolumn{2}{c|}{Method A}
      & \multicolumn{2}{c|}{Method B}\\
      \hline
      $n$ &
      Bead & Cont. &
      Bead & Cont. \\
      \hline 
     
      2   & 3 & 0  & 3 & 0 \\
      4   & 3 & 0  & 3 & 0 \\
      8   & 1 & 0  & 1 & 0 \\
      16  & 1 & 0  & 1 & 0 \\
      32  & 0 & 0  & 0 & 0 \\
      \hline 
    \end{tabular}
    \label{tab:ho-corr-1}
  \end{table}

  \begin{figure}[t]
  \begin{center}
    \includegraphics[scale=1]{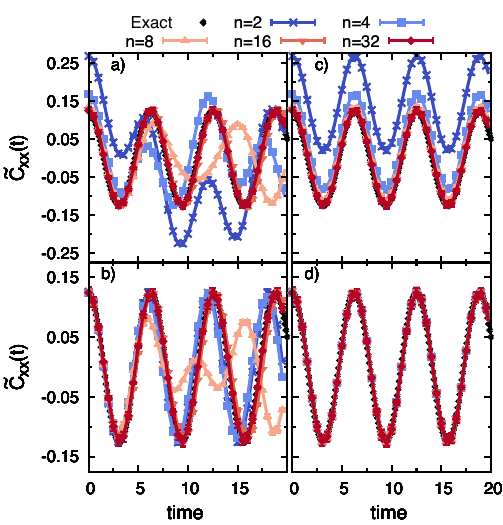}
  \end{center}
  \caption{Kubo-transformed position autocorrelation function for the harmonic oscillator (Eq.~\ref{eq:harm}) at $\beta=8 $ for different BF-RPMD methods using varying numbers of beads, with the values of $k_{\mathrm{max}}$ given in Table~\ref{tab:ho-corr-1}. The top row shows the correlation function calculated with the bead estimator, and the bottom row with the continuous estimator. (a-b) Method A, (c-d) Method B. Exact results are given as black circles.} 
  \label{fig:ho-corr}
  \end{figure}

  We first examine the position autocorrelation function at $\beta=8$. The full set of converged CFs is shown in Fig.~\ref{fig:ho-corr} with the number of Fourier components used for each method and number of beads given in Table~\ref{tab:ho-corr-1}. As with the BF-CMD results in Paper I\cite{octLondon2025}, we need fewer Fourier components to achieve convergence when increasing the number of beads (see Table~\ref{tab:ho-corr-1}). 
  \begin{figure}
  \begin{center}
    \includegraphics[scale=1]{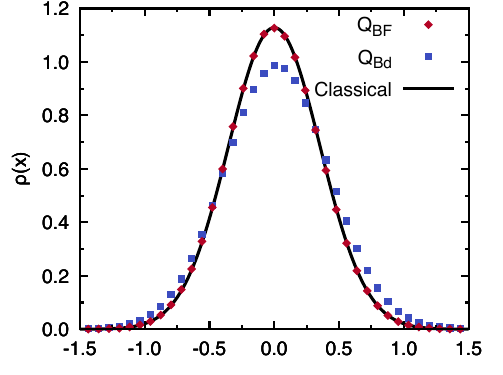}
  \end{center}
  \caption{Position distribution function of the harmonic oscillator (Eq.~\ref{eq:harm}) from BF-RPMD with $n=4$ and $k_{\mathrm{max}}=1 $, comparing the distributions of the bead centroid of Eq.~\ref{eq:bd-cent} (blue squares) and the BF centroid of Eq.~\ref{eq:bf-cent} to the classical Boltzmann distribution (black line).}
  \label{fig:ho-dist-2}
  \end{figure}
When looking at the initial values, the continuous estimator for both Methods A and B (bottom row of Fig.~\ref{fig:ho-corr}) matches the exact result for any number of beads. The bead-estimated values (top row of Fig.~\ref{fig:ho-corr}) only converge to the exact value when using 16 beads. It is known that the zero-time limit of the Kubo-transformed position ACF of a harmonic oscillator is exact even for a classical distribution, and that the centroid of B-PIMD samples the classical Boltzmann distribution.\cite{augCraig2004} When looking at the centroids from BF-PIMD, as shown in Fig.\ref{fig:ho-dist-2}, the BF centroid exactly matches the classical Boltzmann distribution, while the bead centroid does not. The bead centroid sampling the incorrect distribution then gives rise to the 
  incorrect zero-time limit value of the ACF.

The resulting CFs calculated using Method A, i.e., without thermostatting (panels a and b of Fig.~\ref{fig:ho-corr}), exhibit dynamics that are more erratic than pure harmonic motion, particularly when using fewer beads. This non-harmonic motion comes from the coupling between the motion of the Fourier components and the external potential. For the bead estimator, the CF is tracking the dynamics of the bead centroid, which is governed by the force on centroid momenta. The EOM for the bead centroid momenta under the harmonic potential is,
  \begin{equation}
    \begin{aligned}
      \frac{\partial {P}_{\mathrm{bd}}}{\partial t} &= - \frac{1}{n} \sum_{j=1}^n \left[ q_{j} + 
    \mathop{\sum_{k=1}^{k_{\mathrm{max}}}}_{k\in\mathrm{odd}} \frac{2}{k\pi}a_{jk} \right] \\
                                                    &= -Q_{\mathrm{BF}}.
      \end{aligned}
    \label{eq:bd-cent-EOM}
  \end{equation}

Eq.~\ref{eq:bd-cent-EOM} demonstrates that the motion of the bead centroid is not simple harmonic motion, as the force on the bead centroid momentum is the position of the BF centroid instead of the bead centroid. Thus, the motion of the bead centroid is still tied to the motion of the Fourier components. As the number of beads in the RP increases, the distribution of the Fourier components becomes narrower, and the bead centroid and BF centroid become closer in value. As such, the dynamics of the bead centroid approaches simple harmonic motion. This is seen in Fig.~\ref{fig:ho-corr}a, where, as the number of beads increases, the CF reaches the exact result.

The scaling of the Fourier components alters the Hamiltonian such that the frequency of the pure Fourier harmonics is just $\omega_{n}$ instead of $\omega_{n} \frac{(k\pi)^2}{2}$. This change in the frequency also explains the very different behavior for $n=8$, as for that number of beads at $\beta=8$, $\omega_{n}$ becomes one, which matches the frequency of the external potential, creating a resonant behavior.

While the continuous estimator gives improved results for the zero-time limit, the dynamics of the BF centroid are still troubled by couplings between the Fourier components and the external potential, as seen in Fig. \ref{fig:ho-corr}b. Method B, which introduces a thermostat to the dynamics of the Fourier components, removes the resonances between the Fourier components, beads, and the external potential and yields pure harmonic motion. This removal of resonances is similar to that seen for vibrational spectra with T-RPMD.\cite{rossi_trpmd}. The bead-estimated CFs are still limited in their accuracy by the zero-time limit, needing 16 beads to reach the exact value. The continuous estimator gives the exact result with 2 beads, the minimum considered here.
 
\begin{figure}[t]
  \begin{center}
    \includegraphics[scale=1]{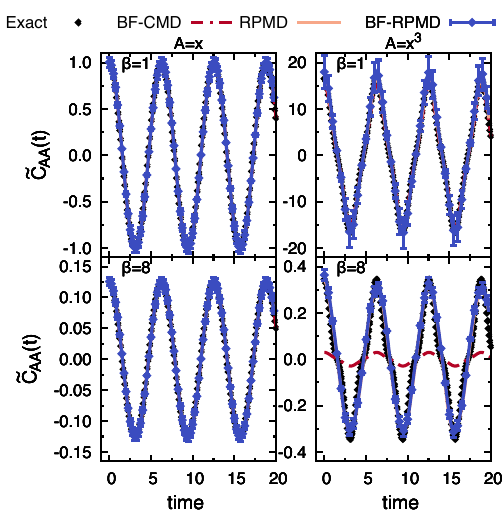}
  \end{center}
  \caption{Kubo-transformed autocorrelation functions for the harmonic oscillator (Eq.~\ref{eq:harm}) at high and low temperatures and for linear and cubic position operators. All BF-RPMD results (shown in blue diamonds) use Method B. Results for $A=x$ use the continuous estimator while results for $A=x^3$ use the bead estimator. 
    $n$;$k_{\mathrm{max}}$ combinations for
$\beta=1$; $A=x$: 2; 0,
$\beta=1$; $A=x^3$: 2; 1,
$\beta=8$; $A=x$: 2; 0,
$\beta=8$; $A=x^3$: 16; 1. BF-CMD results (shown in red dashed-dotted lines) use $n=2$; $k_{\mathrm{max}}=0$. RPMD results (shown in solid orange lines) use $n=4\beta$. Exact results are shown as black circles.}
  \label{fig:ho-corr-comp}
  \end{figure}

These results indicate the need for the thermostatting of the Fourier components. As such, all results shown from here will only include Method B. Additionally, only the converged results will be provided, with full convergence details in the Supplementary Material.

  To end our discussion of the harmonic oscillator, we do a full comparison between BF-RPMD, B-RPMD, BF-CMD, and the exact results for both $\beta=1$ and 8 and the linear and non-linear operators (Fig. \ref{fig:ho-corr-comp}). All methods are in perfect agreement for the linear operator as expected. They also agree for $A=x^3$ in the high-temperature $\beta=1$ case. BF-RPMD is able to converge using 2 beads with one Fourier component, a slight advantage over B-RPMD. The low-temperature case highlights the issues of using BF-CMD for non-linear operators without any corrections.\cite{octLondon2025} BF-RPMD is able to match the exact (and B-RPMD) results using 16 beads and one Fourier component. While a small advantage over B-RPMD in terms of beads, it does showcase that BF-RPMD is able to capture the NQEs in these non-linear operators without any corrections to the correlation function.

  \subsection{Anharmonic Oscillators}
 
  \begin{table}
    \caption{Convergence parameters for the BF-RPMD calculated Kubo-transformed autocorrelation functions of the mildly anharmonic oscillator (Eq.~\ref{eq:mild}).}
    \begin{tabular}{|c | c |  c|}
    \hline
      & $A=x$
      & $A=x^3$ \\
      \hline 
      $\beta$ &
      $n$; $k_{\mathrm{max}}$; Est. &
      $n$; $k_{\mathrm{max}}$; Est. \\
      \hline \hline 
      1 & 2; 0; Cont. & 2; 3; Bead \\
      8 & 4; 3; Cont. & 16; 2; Bead \\
      \hline
    \end{tabular}
    \label{tab:mild-conv}
  \end{table}
  
  \begin{figure}[t]
  \begin{center}
    \includegraphics[scale=1]{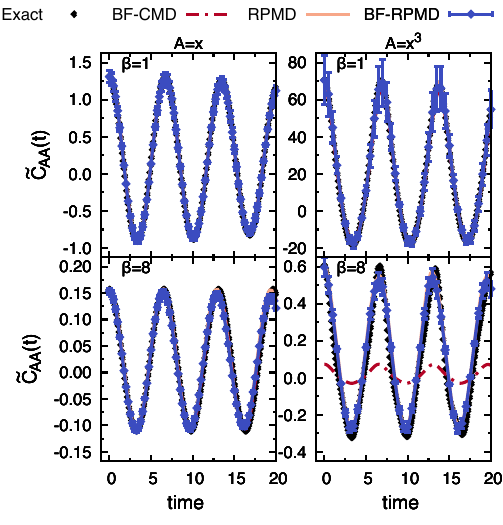}
  \caption{Kubo-transformed autocorrelation functions for the mildly anharmonic oscillator (Eq.~\ref{eq:mild}) at high and low temperatures and for linear and cubic position operators. The convergence parameters for the BF-RPMD results (shown in blue diamonds) are given in Table~\ref{tab:mild-conv}. BF-CMD results (shown in red dash-dotted lines) use $n=2$; $k_{\mathrm{max}}=0$ at $\beta=1$, and $n=4$; $k_{\mathrm{max}}=1$ for $\beta=8$. RPMD results (shown in solid orange lines) use $n=4\beta$. Exact results are shown as black circles.}
  \label{fig:mild-full}
\end{center}
  \end{figure}

As with the harmonic oscillator, for the mildly anharmonic oscillator of Eq.~\ref{eq:mild}, we compare the BF-RPMD results with B-RPMD, BF-CMD, and the exact results. The full set of correlation functions is plotted in Fig.~\ref{fig:mild-full} with the BF-RPMD convergence data given in Table~\ref{tab:mild-conv}.

As with the harmonic oscillator, we see excellent agreement between BF-RPMD and B-RPMD as well as the exact results. Again, we see a considerable underestimation of the correlation function for BF-CMD at low temperature for the non-linear operator. For BF-RPMD's efficiency, we see an eightfold reduction in the number of beads for the linear operator at low temperature by adding 3 Fourier components. Again, as with the harmonic oscillator, a more modest improvement is observed for the non-linear operator, with a twofold improvement upon adding 2 Fourier components.

  \begin{table}
    \caption{Same as Table~\ref{tab:mild-conv} but for the quartic oscillator (Eq.~\ref{eq:quart})}
    \begin{tabular}{|c | c |  c|}
    \hline
      & $A=x$
      & $A=x^3$ \\
      \hline 
      $\beta$ &
      $n$; $k_{\mathrm{max}}$; Est. &
      $n$; $k_{\mathrm{max}}$; Est. \\
      \hline \hline 
      1 & 2; 2; Cont. & 2; 1; Cont. \\
      8 & 16; 2; Cont. & 16; 2; Cont.\\
      \hline
    \end{tabular}
    \label{tab:quart-conv}
  \end{table}
  
  \begin{figure}[t]
  \begin{center}
    \includegraphics[scale=1]{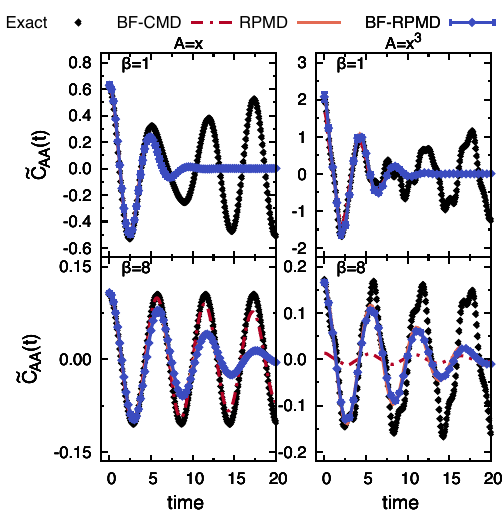}
  \end{center}
  \caption{Kubo-transformed autocorrelation functions for the quartic oscillator (Eq.~\ref{eq:quart}) at high and low temperatures and for linear and cubic position operators. The convergence parameters for the BF-RPMD results (shown in blue diamonds) are given in Table~\ref{tab:quart-conv}. BF-CMD results (shown in red dash-dotted lines) use $n=2$; $k_{\mathrm{max}}=1$ at $\beta=1$, and $n=8$; $k_{\mathrm{max}}=1$ for $\beta=8$. RPMD results (shown in solid orange lines) use $n=4\beta$. Exact results are shown as black circles.}
  \label{fig:quart-full}
  \end{figure}

  The quartic oscillator provides a much greater challenge for all of the methods considered here due to their lack of quantum coherence. Calculated autocorrelation functions for this system are shown in Fig.~\ref{fig:quart-full} with the BF-RPMD convergence parameters given in Table~\ref{tab:quart-conv}. As with the other two systems, BF-RPMD closely matches B-RPMD. The major difference for this system is that it takes 16 beads and 2 Fourier components to reach convergence for both the linear and non-linear operators at $\beta=8$. However, as seen in Fig. S5 of the Supplementary Material, the short-time behavior is captured well with just 8 beads. Given that this short-time limit is where approximate real-time PI methods are most useful, the gains of BF-RPMD are more prevalent than at first glance.

  For the non-linear operator results shown here, we use the continuous estimator. This is due to their improved values for the zero-time limit compared to the bead-estimated values. This indicates that the performance of the estimators could be tied to both the operator and the system. As such, both should be considered when studying such anharmonic systems. From a practical standpoint, however, this does not introduce significant complexity to the simulations, as both can easily be calculated from the same trajectories.

  One drawback of the current BF-RPMD method and implementation is the integrals over the paths needed to calculate the forces and energies. As previously mentioned, these integrals are performed using the trapezoid rule and require many segments to compute. Due to the simplicity of these systems, the overall computational cost is minimal, but could easily become intractable for large-scale atomistic simulations where the force calculations are the computational bottleneck. For those systems, any savings in reducing the number of beads are going to be lost due to an increased number of force evaluations compared to B-RPMD. 
  In Paper III, we will explore an alternate approach to the integrals over the path to make the BF-PIMD/RPMD methods tractable for use in simulations of atomistic systems.
  %\textcolor{purple}{Any future exploration of the method will have to involve improving the computation of those integrals in order for it to be an attractive method for approximate real-time quantum dynamics. However, the method's current ability to provide accuracy on par with B-RPMD with fewer beads without any corrections to the correlation functions makes working to improve the method's efficiency worthwhile.} \textcolor{red}{Revise the above statement to connect to Paper III...}

\section{Conclusions}\label{sec:conc}
In this work, we have presented a novel method for utilizing BF-PIs for approximate real-time quantum dynamics through the development of the BF-RPMD method. Through the study of autocorrelation functions for a number of model systems, we determined the best formulation of BF-RPMD in which a scaling is applied to the amplitudes of the Fourier components and a thermostat to their momenta during the dynamics. The convergence of BF-RPMD with respect to the number of beads and Fourier components is highly system-dependent as well as operator-dependent. For linear operators and mildly anharmonic systems, up to an eightfold reduction in the number of beads is obtained compared to RPMD when 3 Fourier components are included, achieving performance similar to BF-CMD. For the quartic model, a less significant, twofold reduction in beads is achieved when adding 2 Fourier components. Similarly, a twofold reduction in beads is observed across all systems when computing the CF for the cubic position operator. However, the overall performance of BF-RPMD for non-linear operators is vastly superior to that of BF-CMD.

The results for BF-RPMD shown here show promise and beckon further development. An obvious direction to pursue in the future is to study the performance of BF-PIMD and BF-RPMD methods for more complex systems, particularly large-scale condensed-phase atomistic systems. To do so, in Paper III, we present the first implementation of the BF-PIMD and BF-RPMD methods for atomistic simulations, providing practical recipes for additional force calculations along the BF paths.

In our future studies, we will also explore the types of dynamical properties that BF-RPMD is best suited to calculate. In particular, we are interested in how changes in the frequency of the internal RP modes due to reducing the number of beads, and changes in the external potential from the BF paths, affect the spurious peaks that appear in RPMD vibrational spectra. As part of that study, we shall examine the fully thermostatted BF-RPMD method. Additionally, we shall explore combining the BF-RPMD method with surface-hopping\cite{Shakib2017RingPolymer, Limbu2023RealTime, Brown2021RecentProgress, Limbu2025Diagonal} for efficient non-adiabatic quantum dynamics simulations.

\section*{Supplementary Material}
Convergence of Fourier components for the harmonic, mildly anharmonic, and quartic oscillators and the calculated correlation functions.

\begin{acknowledgments}
Simulations presented in this work used resources from Bridges-2 at Pittsburgh Supercomputing Center through allocations PHY230030P and CHE240103 from the Advanced Cyberinfrastructure Coordination Ecosystem: Services \& Support (ACCESS) program,\cite{access} which is supported by National Science Foundation grants \#2138259, \#2138286, \#2138307, \#2137603, and \#2138296. The authors also acknowledge the HPC center at UMKC for providing computing resources and support.
\end{acknowledgments}

\subsection*{Conflict of Interest}

\noindent The authors have no conflicts to disclose.

\subsection*{Author Contributions}
\textbf{Nathan London}: Method (equal); Writing - original draft (lead); Writing - review \& editing (equal). 
\textbf{Mohammad R. Momeni}: Conceptualization (lead); Funding acquisition (lead); Supervision (lead); Method (equal); Writing - original draft (supporting); Writing - review \& editing (equal). 

\section*{\label{sec_contrib} DATA AVAILABILITY}
\noindent  The data that support the findings of this study are provided in the main text and the Supplementary Material. Additional data are available from the corresponding author upon request.

\section*{REFERENCES}
\bibliography{bf-rpmd}
\end{document}